\documentclass[bibyear]{aa}
\usepackage{graphicx}
\usepackage{txfonts}
\newcommand{\lpup}{\mbox{L$_2$~Pup}}

\newcommand{\MSOLPERYR}{\mbox{$M_{\sun}$~yr$^{-1}$}}
\newcommand{\micron}{\mbox{$\mu$m}}
\newcommand{\KMS}{\mbox{km s$^{-1}$}}
\newcommand{\PHISTAR}{\mbox{$\phi_{\rm \star}$}}
\newcommand{\SIGMAEW}{\mbox{$\sigma_{\rm major}$}}
\newcommand{\SIGMANS}{\mbox{$\sigma_{\rm minor}$}}
\newcommand{\FSTAR}{\mbox{$f_{\star}$}}
\newcommand{\PERRAD}{\mbox{rad$^{-1}$}}
\newcommand{\RRINGEW}{\mbox{$R_{\rm major}$}}
\newcommand{\RRINGNS}{\mbox{$R_{\rm minor}$}}
\usepackage{color}

\begin{document}

\title{
AMBER-NACO aperture-synthesis imaging of the half-obscured central star 
and the edge-on disk of the red giant \lpup
\thanks{
Based on AMBER, NACO, and MIDI observations made with the Very Large Telescope 
and Very Large Telescope Interferometer of the European Southern Observatory. 
Program ID: 074.D-0075(A), 074.D-0101(A), 074.D-0198(B), 088.D-0150(A/B), 
and 288.D-5041(A)
}
}
%\subtitle{
%}

\author{K.~Ohnaka\inst{1,2} 
\and
D.~Schertl\inst{2}
\and
K.-H.~Hofmann\inst{2} 
\and
G.~Weigelt\inst{2} 
}

\offprints{K.~Ohnaka}

\institute{
Universidad Cat\'{o}lica del Norte, Instituto de Astronom\'{i}a, 
Avenida Angamos 0610, Antofagasta, Chile\\
\email{k1.ohnaka@gmail.com}
\and
Max-Planck-Institut f\"{u}r Radioastronomie, 
Auf dem H\"{u}gel 69, 53121 Bonn, Germany
}

\date{Received / Accepted }

\abstract
{}
% Aim
{
The red giant \lpup\ started a dimming event in 1994, which is considered to 
be caused by the ejection of dust clouds.  
We present near-IR aperture-synthesis imaging of \lpup\ 
achieved by combining data from VLT/NACO and the AMBER instrument of the Very Large Telescope Interferometer (VLTI). Our aim is to 
spatially resolve the innermost region of the circumstellar environment. 
}
% Methods
{
We carried out speckle interferometric observations at 2.27~\micron\ 
with VLT/NACO and long-baseline interferometric observations with VLTI/AMBER 
at 2.2--2.35~\micron\ with baselines of 15--81~m. 
We also extracted an 8.7~\micron\ image from the mid-IR VLTI 
instrument MIDI. 
}
% Results
{
The diffraction-limited image obtained by bispectrum speckle interferometry 
with NACO 
with a spatial resolution of 57~mas shows an elongated component. 
The aperture-synthesis imaging combining the NACO speckle data and AMBER data 
with a spatial resolution of 5.6$\times$7.3~mas further resolves 
not only this elongated component, but also the central star. 
The reconstructed image reveals that the elongated component is a nearly 
edge-on disk with 
a size of $\sim \!\! 180\times50$~mas lying in the E-W direction, 
and furthermore, that the southern hemisphere of the central star is severely 
obscured by the equatorial dust lane of the disk. 
The angular size of the disk is consistent with the distance that 
the dust clouds that were ejected at the onset of the dimming event should have 
traveled by the time of our observations, if we assume that the dust clouds 
moved radially. 
This implies that the formation of the disk may be responsible for the 
dimming event. 
The 8.7~\micron\ image with a spatial resolution of 220~mas extracted 
from the MIDI data taken in 2004 
(seven years before the AMBER and NACO observations) 
shows an approximately spherical envelope without a signature of the disk. 
This suggests that the mass loss before the dimming event 
may have been spherical. 
}
% Conclusions
{}

   %  END OF ABSTRACT

\titlerunning{AMBER-NACO imaging of the half-obscured central star and 
  the edge-on disk of \lpup}
\authorrunning{Ohnaka et al.}
\maketitle

\section{Introduction}
\label{sect_intro}

Slow, but intense mass loss at the asymptotic giant branch (AGB) 
leads to the formation of thick circumstellar envelopes.   
When AGB stars evolve further to protoplanetary nebulae (PPNe) and 
planetary nebulae (PNe), striking bipolar lobes emerge, often accompanied 
by collimated jets (Sahai \& Trauger \cite{sahai98}). 
Binarity is currently considered to be the most 
promising mechanism to shape PNe (e.g., De Marco \cite{demarco09}), 
but it is not yet understood how and at which evolutionary phase the bipolar 
structure emerges.  
High-resolution imaging of a few dusty AGB stars 
reveals bipolar or more complex, clumpy structures (e.g., 
Weigelt et al. \cite{weigelt98}; Monnier et al. \cite{monnier00}; 
Hofmann et al. \cite{hofmann01}).  
Some AGB stars show spiral structures, which strongly suggests the 
influence of a companion 
(e.g., Mauron \& Huggins \cite{mauron06}; 
Maercker et al. \cite{maercker12}; Kim et al. \cite{kim13}; 
Ramstedt et al. \cite{ramstedt14}; Mayer et al. \cite{mayer14}; 
Decin et al. \cite{decin15}). 
These observations suggest that the seed of the morphological change 
already exists in the AGB phase.

The M5 giant L$_2$ Pup is a bright, nearby AGB star 
with a mass-loss rate of 
$\sim \!\! 3 \times 10^{-7}$ \MSOLPERYR\ (Jura et al. \cite{jura02}) that 
shows episodic, asymmetric dust formation.  
\mbox{Magalh\~aes} et al. (\cite{magalhaes86}) interpreted temporally variable 
polarization in the optical as due to the growth and 
dissipation of dust grains in an asymmetric dust cloud. 
The 11.7 and 17.9~\micron\ images of Jura et al. (\cite{jura02}) 
showed an asymmetric circumstellar dust envelope.  
Furthermore, Bedding et al. (\cite{bedding02}) have detected a dimming 
event in the optical light curve of L$_2$ Pup starting in 1994, which 
suggests the ejection of dust clouds.  
The brightness profiles reconstructed in the visible also lend support 
to the hypothesis that there are asymmetric dust clouds close to the star 
(Ireland et al. \cite{ireland04}).  

The circumstellar environment of \lpup\ close to the star has recently been 
investigated in more detail. 
Kervella et al. (\cite{kervella14}, hereafter K14) observed \lpup\ using 
the adaptive optics instrument NACO at VLT. 
Their images taken from 1.04 to 4.05~\micron\ with a spatial resolution 
from 26 to 100~mas reveal a nearly edge-on disk lying in 
the E-W direction.  
The authors also showed that the observed images can be explained well by a flared 
disk model. 
The 2.30 and 3.74~\micron\ images obtained by Lykou et al. (\cite{lykou15}, 
hereafter L15) 
using the aperture-masking technique with NACO 
also show an elongated component.  
More recently, Kervella et al. (\cite{kervella15}) have revealed a 
clear bipolar structure using the VLT/SPHERE-ZIMPOL instrument. 

In this paper, we present near-IR interferometric observations 
of \lpup\ with an even higher spatial resolution of 5.6$\times$7.3~mas, 
combining VLT/NACO and the near-IR interferometric instrument AMBER 
at the Very Large Telescope Interferometer (VLTI).  
In addition, we also present a mid-IR image at 8.7~\micron\ extracted from the 
mid-IR VLTI instrument MIDI.

\begin{figure}
\begin{center}
\resizebox{7cm}{!}{\rotatebox{-90}{\includegraphics{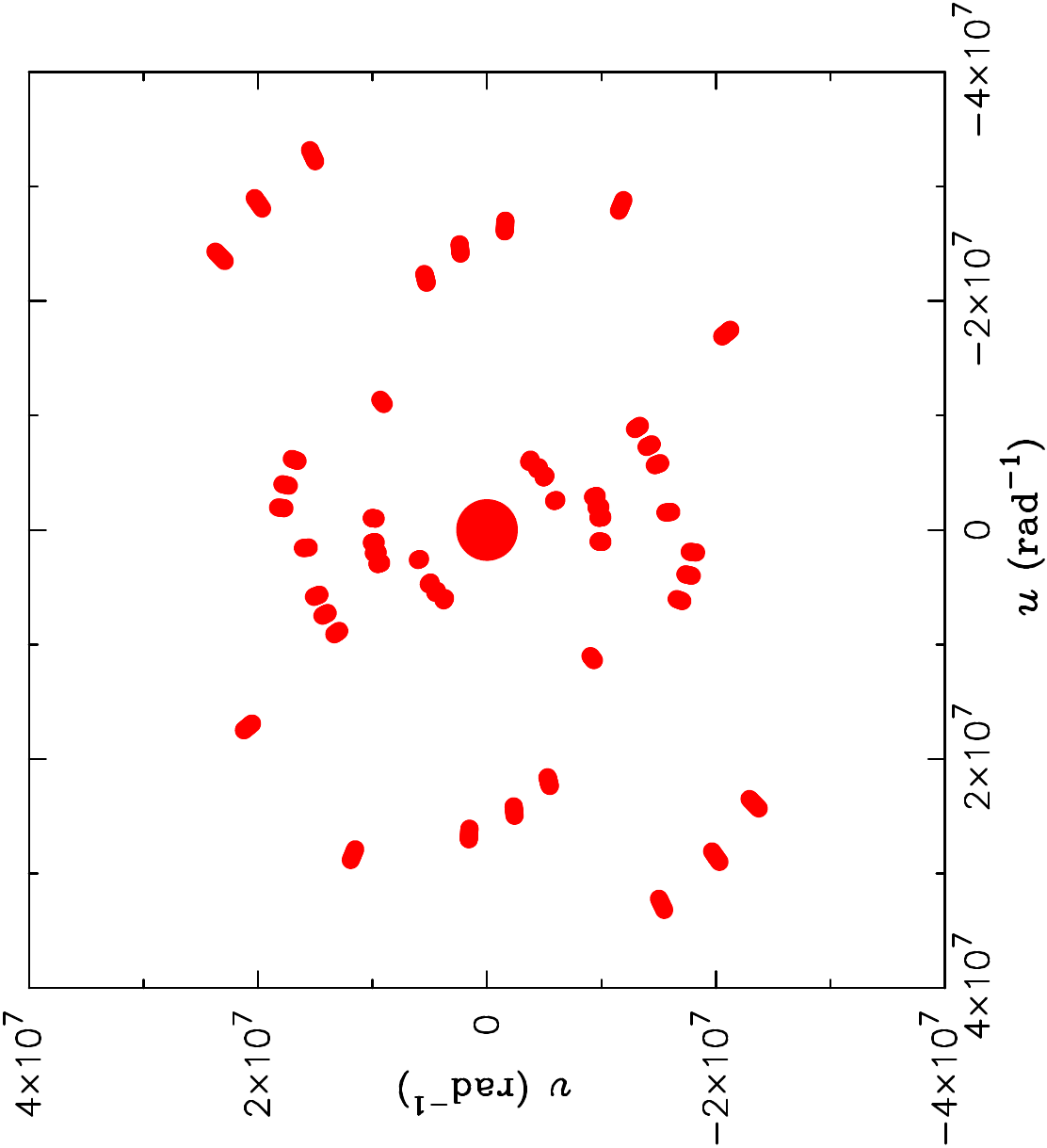}}}
\end{center}
\caption{
$u\varv$ coverage of our AMBER and NACO speckle observations of \lpup.  
}
\label{uvcoverage}
\end{figure}

\begin{table*}
\begin{center}
\caption {
Summary of the AMBER and NACO speckle observations and the archived MIDI data 
of \lpup. 
The seeing, which was measured with a differential image motion monitor 
(DIMM), and the coherence time ($\tau_0$) were measured in the visible 
(see for details 
http://www.eso.org/gen-fac/pubs/astclim/paranal/seeing/adaptive-optics/).  
N$_{\rm f}$ is the number of frames, and N$_{\rm exp}$ is the number of exposures.
}
\vspace*{-2mm}

\begin{tabular}{l c c c c r l l}\hline
\# & $t_{\rm obs}$ & $B_{\rm p}$ & PA     & Seeing   & $\tau_0$ &
${\rm DIT}\times{\rm N}_{\rm f}\times{\rm N}_{\rm exp}$ & Remarks\\ 
& (UTC)       & (m)       & (\degr) & (\arcsec)       &  (ms)    &  (ms) & \\
\hline
\multicolumn{8}{c}{AMBER (A1-C1-D0): \lpup : 2011 Dec 23 (UTC)}\\
\hline
1 & 01:48:07 & 35.7/14.5/22.3 & $-175$/23/174    & 0.9 & 4.0 & $100\times500\times5$& \\
2 & 02:33:57 & 35.7/14.9/22.3 & $-166$/34/$-180$ & 1.7 & 2.5 &
$100\times500\times5$&Not used \\
3 & 03:11:28 & 35.8/15.3/22.3 & $-160$/42/$-174$ & 1.2 & 3.0 & $100\times500\times5$& \\
4 & 03:47:56 & 35.8/15.6/22.2 & $-153$/50/$-169$ & 0.9 & 3.7 & $100\times500\times5$& \\
5 & 04:24:20 & 35.7/15.9/22.1 & $-147$/57/$-164$ & 1.1 & 3.5 & $100\times500\times10$& \\
6 & 06:00:50 & 34.4/15.8/21.4 & $-132$/74/$-151$ & 1.7 & 1.7 &
$100\times500\times10$&Not used \\
7 & 07:04:38 & 32.5/15.1/20.5 & $-123$/85/$-144$ & 2.2 & 1.3 &
$100\times500\times15$&Not used \\
8 & 07:23:33 & 31.8/14.7/20.2 & $-121$/88/$-142$ & 2.2 & 1.4 &
$100\times500\times10$&Not used \\
9 & 08:41:54 & 27.4/12.8/18.2 & $-111$/103/$-134$& 2.7 & 1.1 &
$100\times500\times5$&Not used \\
C1 & 01:19:52& 30.2/10.3/21.5 & $-174$/33/173      & 0.8& 4.7 & $100\times1000\times 5$&Sirius\\
C2 & 02:16:06& 31.0/12.0/21.5 & $-162$/47.5/$-178$ & 0.9& 4.1 & $100\times500\times5$&Sirius\\
C3 & 03:30:29& 33.1/14.3/21.7 & $-149$/60/$-167$   & 0.9& 4.1 & $100\times500\times5$&Sirius\\
C4 & 04:06:45& 34.2/15.1/22.0 & $-144$/64/$-162$   & 0.8& 4.7 & $100\times500\times5$&Sirius\\
C5 & 04:50:11& 35.2/15.8/22.3 & $-138$/69/$-157$   & 1.3& 2.7 & $100\times500\times10$&Sirius\\
\hline
\multicolumn{8}{c}{AMBER (D0-H0-I1): 2011 Dec 29 (UTC)}\\
\hline
10&06:46:01 & 81.0/40.0/60.0 & 114/$-21$/86     & 0.9 & 7.0 & $100\times500\times8$& \\
11&07:38:25 & 78.1/40.4/55.6 & 124/$-13$/95     & 0.9 & 6.0 & $100\times500\times5$& \\
12&08:20:26 & 75.1/40.5/51.0 & 134/$-67$/103    & 0.9 & 6.5 & $100\times500\times5$& \\
13&08:56:34 & 72.3/40.5/46.6 & 142/$-104$/111  & 1.0--1.7 & 4.4 &
$100\times500\times5$&Not used\\
C6& 06:23:18& 76.7/39.2/61.1 & 107/$-205$/77    & 1.0 & 5.8 & $100\times500\times5$&Sirius\\
C7& 07:10:38& 69.8/38.5/56.3 & 113/$-137$/79    & 2.0 & 2.8 & $100\times500\times1$&Sirius\\
C8& 07:20:40& 73.1/40.6/49.0 & 136/$-44$/105    & 1.2 & 4.6 & $100\times500\times5$&Sirius\\
\hline
\multicolumn{8}{c}{AMBER (D0-H0-I1): 2011 Dec 30 (UTC)}\\
\hline
14&02:28:00 & 68.4/32.2/60.5 & 67/$-51$/39     & 0.7 & 6.0 & $100\times500\times5$& \\
15&05:23:07 & 82.2/39.0/63.6 & 100/$-32$/72    & 1.5 & 2.7 & $100\times500\times5$&Not
used\\
C9&00:09:57 & 58.5/39.5/42.5 & 95/$-39$/52     & 1.2& 3.8 & $100\times500\times5$&Rigel\\
C10&01:10:28 & 50.7/36.8/43.2 & 83/$-40$/38    & 1.1 & 4.1 & $100\times500\times5$&Sirius\\
C11&01:52:41 & 61.3/38.5/48.6 & 87/$-40$/48    & 0.9 & 4.7 & $100\times500\times5$&Sirius\\
C12&03:47:19 & 81.4/39.0/64.0 & 101/$-29$/73   & 0.6 & 7.6 & $100\times500\times5$&Rigel\\
\hline
\multicolumn{8}{c}{NACO: 2012 Mar 21 (UTC)}\\
\hline
16&02:30:14 & --- & ---     & 1.6 & 3.0 & $109 \times 398 \times 1$& \\
C13& 02:38:11& --- & --- & 2.0 & 3.0 & $109 \times 300 \times 1$& Canopus\\
\hline
\multicolumn{8}{c}{MIDI (UT3-UT4): 2004 Dec 31 (UTC)}\\
\hline
17&03:19:33 & --- & --- & 1.2 & 2.3 & $4\times1000\times2$& Acquisition images\\
18&05:02:14 & --- & --- & 1.7 & 1.5 & $4\times1000\times2$& Acquisition images\\
C14&01:47:07& --- & --- & 1.1 & 2.6 & $4\times1000\times2$& S~Ori\\
C15&02:37:28& --- & --- & 0.8 & 3.6 & $4\times1000\times1$& S~Ori\\
C16&07:09:13& --- & --- & 0.9 & 2.7 & $4\times1000\times2$& $\varepsilon$~Cru\\
\hline
\label{obslog}
\vspace*{-7mm}

\end{tabular}
\end{center}
\end{table*}

\section{Observations}
\label{sect_obs}

The near-IR interferometric instrument VLTI/AMBER combines three 
telescopes and allows us to observe objects with a spatial resolution 
of $\sim$3~mas (at 2~\micron) with the currently available baselines and 
a spectral resolution of 35, 1500, and 12000 (Petrov et al. \cite{petrov07}). 
Our AMBER observations of \lpup\ reported in this paper took place on 
2011 December 23, 29, and 30 (UTC) using two different telescope 
configurations (A1-C1-D0 and D0-H0-I1), 
which provided projected baseline lengths from 15 to 81~m 
(Program ID: 088.D-0150A/B). 
The $u\varv$ coverage is shown in Fig.~\ref{uvcoverage}. 
We used the medium spectral resolution of 1500 in the wavelength region 
between 2.2 and 2.35~\micron, covering the continuum and the CO first 
overtone bands. 
The Detector Integration Time (DIT) was set to 100~ms in all AMBER data sets. 
We observed Sirius ($\alpha$~CMa, A1V, uniform-disk diameter = 5.9~mas; 
Davis et al. \cite{davis11}) and Rigel ($\beta$~Ori, B8Iae, 
uniform-disk diameter = 2.43~mas; Richichi et al. \cite{richichi05}) 
as interferometric calibrators. 

Additionally, we carried out speckle interferometric observations with VLT/NACO 
in Director's Discretionary Time 
to obtain interferometric data at short baselines 
(Program ID: 288.D-5041A).  This is crucial 
for the aperture-synthesis imaging of objects with a very extended component. 
The NACO speckle observations reported here occurred on 2012 March 21 
with the IB2.27 filter, which is centered at 2.27~\micron\ with a FWHM 
of 0.06~\micron. 
We recorded 398 frames with DIT = 109~ms using the S13 camera with a 
pixel scale of 13.2~mas.  
The window size was $512\times514$ pixels, which resulted in a field of 
view of $6\farcs8 \times 6\farcs8$.  
We observed Canopus ($\alpha$~Car, A9II, angular diameter = 6.9~mas; 
Domiciano de Souza et al. \cite{domiciano08}) as a calibrator and 
took 300 frames with the same DIT as \lpup. 

As complementary data, 
we also extracted mid-IR images at 8.7~\micron\ (filter FWHM = 1.75~\micron) 
from the acquisition image data obtained with the VLTI/MIDI instrument 
(Leinert et al. \cite{leinert03}). 
These images were taken to adjust the position of the target on the detector. 
Thanks to the adaptive optics system for the MIDI 
operation with the 8.2~m Unit Telescopes (UTs) and the short DIT of 4~ms, 
the MIDI acquisition images are diffraction limited.  
In the ESO archive, there are MIDI data of \lpup\ taken on 2004 December 31, 
2005 January 3, 2005 February 28, and 2005 March 3 (the interferometric data 
obtained on 2005 January 3 (UTC) are presented in K14).  
The data taken on 2004 December 31 with UT3 and UT4 
(Program ID: 074.D-0198B) provide the best images of \lpup\ and the 
calibrators, which are used as references of the point spread function (PSF), 
and we therefore present only these images in this paper.  
We reduced the data of $\varepsilon$~Cru and S~Ori obtained on the 
same night to ascertain the PSF (Program IDs: 074.D-0101A and 074.D-0075A). 
As we show below (Fig.~\ref{l2pup_midi_acq}), 
the quality of the PSF obtained from the data of 
$\varepsilon$~Cru is noticeably poorer than the image quality of \lpup\ 
because the former star is much fainter ($F_{12~\micron} = 32.4$~Jy) than 
the latter ($F_{12~\micron} = 2415$~Jy).  As a second PSF, 
we therefore used the Mira star S~Ori because its mid-IR flux 
($F_{12~\micron} = 151$~Jy) is much higher than $\varepsilon$~Cru. 
Although S~Ori was observed as a science target of interferometric 
observations, it is unresolved in the acquisition images at 8.7~\micron,\ 
as we show below. 
The summary of our AMBER and NACO observations as well as the MIDI 
observations is given in Table.~\ref{obslog}.

We reduced the AMBER data with the amdlib ver.~3.0.3 
package\footnote{http://www.jmmc.fr/data\_processing\_amber.htm}, 
which is based on the P2VM 
algorithm (Tatulli et al. \cite{tatulli07}).  The product of the reduction 
is the visibility (or visibility amplitude), closure phase, 
and differential phase. 
We binned the raw data in the spectral direction with a box car 
filter with a width of three pixels to increase the S/N ratio. 
The details of the reduction are described in 
Ohnaka et al. (\cite{ohnaka09}). 
We excluded the data sets 
\#2, \#6--9, \#13, and \#15.  They were 
taken under poor and variable seeing ($\ga$1\farcs5) and/or short coherence time 
($\la$3.0~ms), which makes the absolute calibration of the visibilities 
unreliable. 
The NACO data were reduced with the bispectrum speckle interferometry method 
(Weigelt \cite{weigelt77}; Lohmann et al. \cite{lohmann83}; 
Hofmann \& Weigelt \cite{hofmann86}).  
The result of the reduction of speckle data is the 2D visibility, 
the bispectrum on a huge number 
of baseline triangles within the maximum baseline of 8~m (aperture of the 
telescope), and a reconstructed image of the object. 
For the combined image reconstruction with AMBER and NACO, 
we sampled visibilities and closure phases from the speckle data 
on randomly selected 855 baseline triangles within a maximum baseline length 
of 6~m (we avoided sampling between 6 and 8~m because the data are noisy). 

For the reconstruction of an aperture-synthesis image from the 
combined AMBER and NACO speckle data, 
we used the image reconstruction package 
MiRA\footnote{http://cral.univ-lyon1.fr/labo/perso/eric.thiebaut/?Software/MiRA} 
(Thi\'ebaut \cite{thiebaut08}), 
which was also used for the image reconstruction of the red supergiants 
Betelgeuse and Antares in our previous works 
(Ohnaka et al. \cite{ohnaka11}, \cite{ohnaka13}).  
Given that the wavelengths covered by AMBER and NACO overlap 
(2.2--2.35 and 2.27~\micron, respectively), it can be justified 
to combine the AMBER and NACO data for the image reconstruction 
in the continuum.  
We used all wavelengths in the continuum region (2.2--2.29~\micron) 
in the AMBER data to reconstruct one image 
because the object image is not expected to change noticeably 
in this narrow continuum region. 
Details of the image reconstruction procedure we adopted are described in 
Appendix~\ref{appendix_reconst}.

The reduction of the MIDI acquisition images is similar to the processing 
described in Ohnaka (\cite{ohnaka14}).  After the sky subtraction, 
we recentered each frame and added the frames.  
The sky-subtracted images of \lpup\ show 
detector artifacts that are presumably caused by the high brightness of \lpup. 
After removing these artifacts as described in Appendix~\ref{appendix_midi}, 
we added all frames in each data set of \lpup.  The calibrators are 
much fainter than \lpup, and the residual of the sky subtraction is 
significant in some frames.  Therefore, we only added the frames in which 
the residual of the sky subtraction is small, to obtain as clean a PSF 
as possible.  
The resulting image from each data set was derotated so that North is up 
and East is to the left, using the field rotation angle 
(Mathar \cite{mathar06}).  Finally, the derotated images from all data sets 
were added.  
The quality of the images taken with UT3 is significantly lower than 
that taken with UT4, therefore we present the final image obtained 
from the UT4 data. 
We carried out neither the PSF subtraction nor the deconvolution because 
the images of \lpup\ are saturated at the central peak.

\section{Results}
\label{sect_res}

\subsection{Near-IR AMBER--NACO aperture-synthesis image}
\label{subsect_res_NIR}

\begin{figure*}
\resizebox{\hsize}{!}{\rotatebox{-90}{\includegraphics{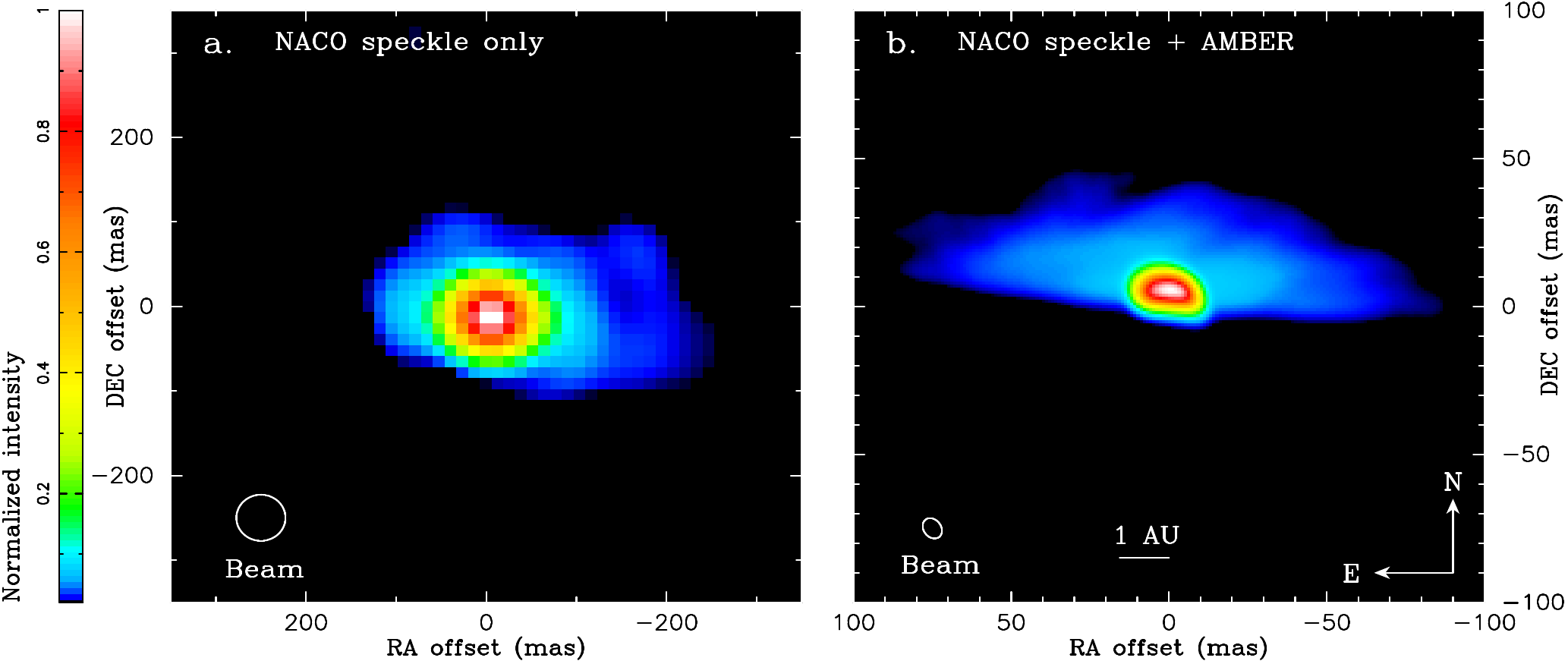}}}
\caption{
{\bf a:} Image of \lpup\ reconstructed from the NACO speckle data alone. 
{\bf b:} Image reconstructed from the combined AMBER and NACO speckle data. 
}
\label{reconst_images}
\end{figure*}

\begin{figure*}
\resizebox{\hsize}{!}{\rotatebox{-90}{\includegraphics{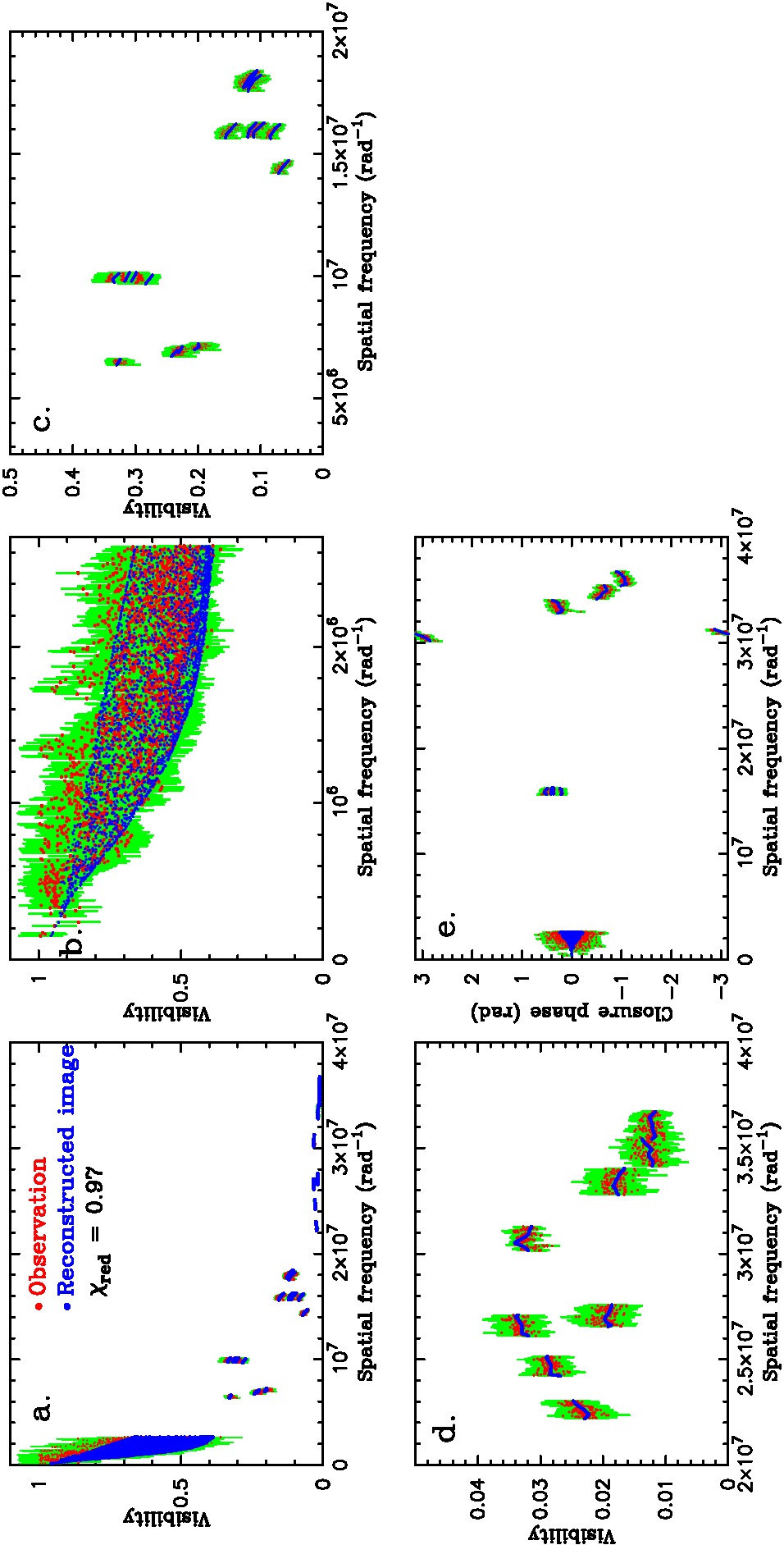}}}
\caption{
Comparison of the observed interferometric data and the corresponding 
observables from the reconstructed image shown in Fig.~\ref{reconst_images}b 
(a color version of the figure is available in the electronic edition).  
The observed data are plotted 
with the red (black in the printed edition) 
symbols with the error bars shown in green (light gray), while the model is 
represented by the blue (dark gray) symbols. 
{\bf a:} Visibilities. 
{\bf b:} Enlarged view of the visibilities obtained by 
speckle interferometry with NACO. 
{\bf c} and {\bf d:} Enlarged views of the visibilities obtained with AMBER 
in two different spatial frequency ranges. 
{\bf e:} Closure phases. 
}
\label{reconst_interf}
\end{figure*}

\begin{figure*}
\sidecaption
\resizebox{12cm}{!}{\rotatebox{0}{\includegraphics{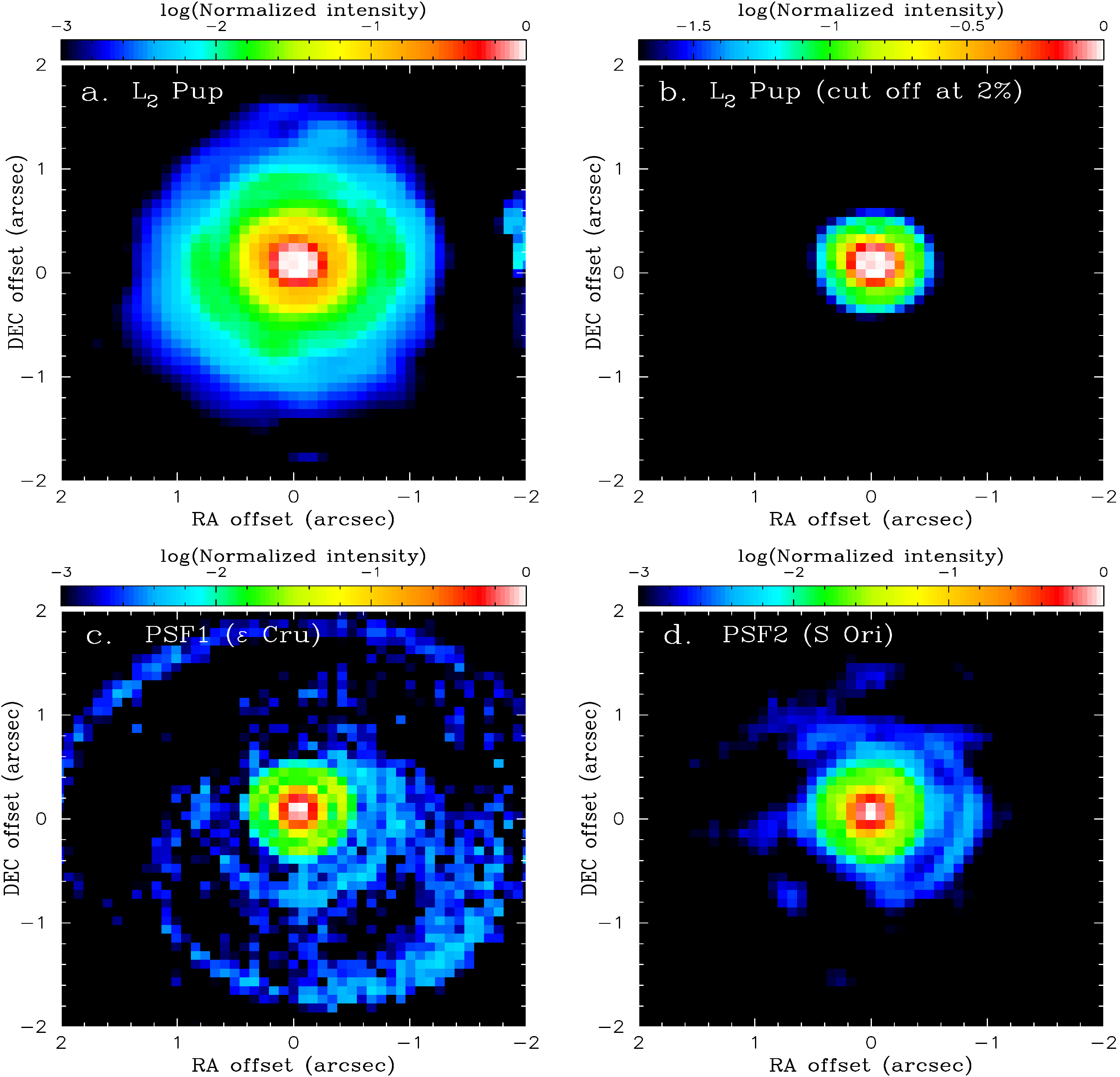}}}
\caption{
8.7~\micron\ images of \lpup\ and two PSF reference stars extracted from 
the MIDI acquisition data.  In all panels, North is up and East is to the 
left. 
{\bf a:} \lpup\ with a minimum intensity of 0.1\% of the central peak. 
{\bf b:} \lpup\ with a minimum intensity of 2\% of the central peak to 
compare with the 11.7~\micron\ image obtained by Jura et al. (\cite{jura02}). 
{\bf c:} PSF reference star $\varepsilon$~Cru.  The large arcs near the top 
and the bottom of the image are the residual of the background subtraction. 
{\bf d:} PSF reference star S~Ori.  
}
\label{l2pup_midi_acq}
\end{figure*}

Figure~\ref{reconst_images} shows the images reconstructed 
from the NACO speckle data alone 
and from the combined NACO speckle and AMBER data. 
Both images are convolved with the following Gaussian beam. 
The beam size (FWHM) of the image reconstructed from the speckle data alone 
is 57~mas, which corresponds to the diffraction-limit of VLT at 2.27~\micron. 
The beam size of the image reconstructed from the combined data is 
$5.6 \times 7.3$~mas, which was derived by fitting the central peak of the 
dirty beam with a \mbox{2D} elliptical Gaussian. 
Figure~\ref{reconst_interf} shows a comparison of the visibilities 
and closure phases from the reconstructed image with the observed data. 
The reduced $\chi^2$ of the fit is 0.97.  
The image reconstructed from the speckle data alone shows a central star 
and an extended component elongated approximately in the E-W direction with 
a size of $\sim \!\! 300\times 200$~mas.  
The size of the extended component in the deconvolved images obtained 
by K14 and L15 in the $K$ band is 
$\sim \!\! 200 \times 100$~mas, which roughly agrees with the measured 
size in our speckle image, which is convolved with the aforementioned beam. 

The image reconstructed from the combined NACO speckle and AMBER data, shown 
in Fig.~\ref{reconst_images}b, further resolves not only the elongated 
component seen in the speckle image, but also the central star. 
In this image, the extended component has a size of 
$\sim \!\!$ 180 $\times$ 50~mas.  
We checked that the NACO speckle + AMBER image matches the NACO speckle-only 
image when convolved with the corresponding beam. 
Moreover, the figure shows that the southern half of the central star 
is severely obscured. 
This AMBER--NACO aperture-synthesis image directly shows both the edge-on 
disk and the half-obscured central star due to the equatorial dust lane 
of the disk. 
At the distance of $64 \pm 4$~pc (from the Hipparcos parallax of 
$15.61 \pm 0.99$~mas, van Leeuwen \cite{vanleeuwen07}), the imaged size of 
the disk corresponds to $(11.5 \pm 0.8) \times (2.8 \pm 0.2)$~AU.

Our combined aperture-synthesis image is qualitatively consistent with 
the disk model of K14. 
They also found out that the 2--2.4~\micron\ interferometric data of \lpup\ 
taken in 2001 by the VLTI/VINCI instrument are better fitted with an ellipse 
with an axis ratio 
of 1.5 with the major axis at a position angle 106\degr\ (i.e., nearly 
in the E-W direction). 
This is naturally explained by the obscuration of the southern half of the 
central star by the edge-on disk, which makes the star appear to be elongated 
in the E-W direction. 

Our AMBER data cover the wavelength region of the CO first overtone bands. 
However, the speckle data were taken only in the continuum 
at 2.27~\micron, not in the CO bands.  This makes the image reconstruction 
more difficult, if not impossible, in the CO bands. 
Therefore, the analysis of the AMBER 
data in the CO bands will be presented in a separate, forthcoming paper. 

We note that the obscuration of the central star by a dust disk was also imaged 
for the eclipsing binary $\varepsilon$~Aur in the $H$ band (1.50--1.74~\micron) 
by Kloppenborg et al. (\cite{kloppenborg10}).  In contrast to \lpup, however, 
the disk of $\varepsilon$~Aur is located around an orbiting companion, 
and the disk itself is not visible at least in the $H$ band.

\subsection{Mid-IR MIDI acquisition image}
\label{subsect_res_MIR}

\begin{figure}
\resizebox{\hsize}{!}{\rotatebox{-90}{\includegraphics{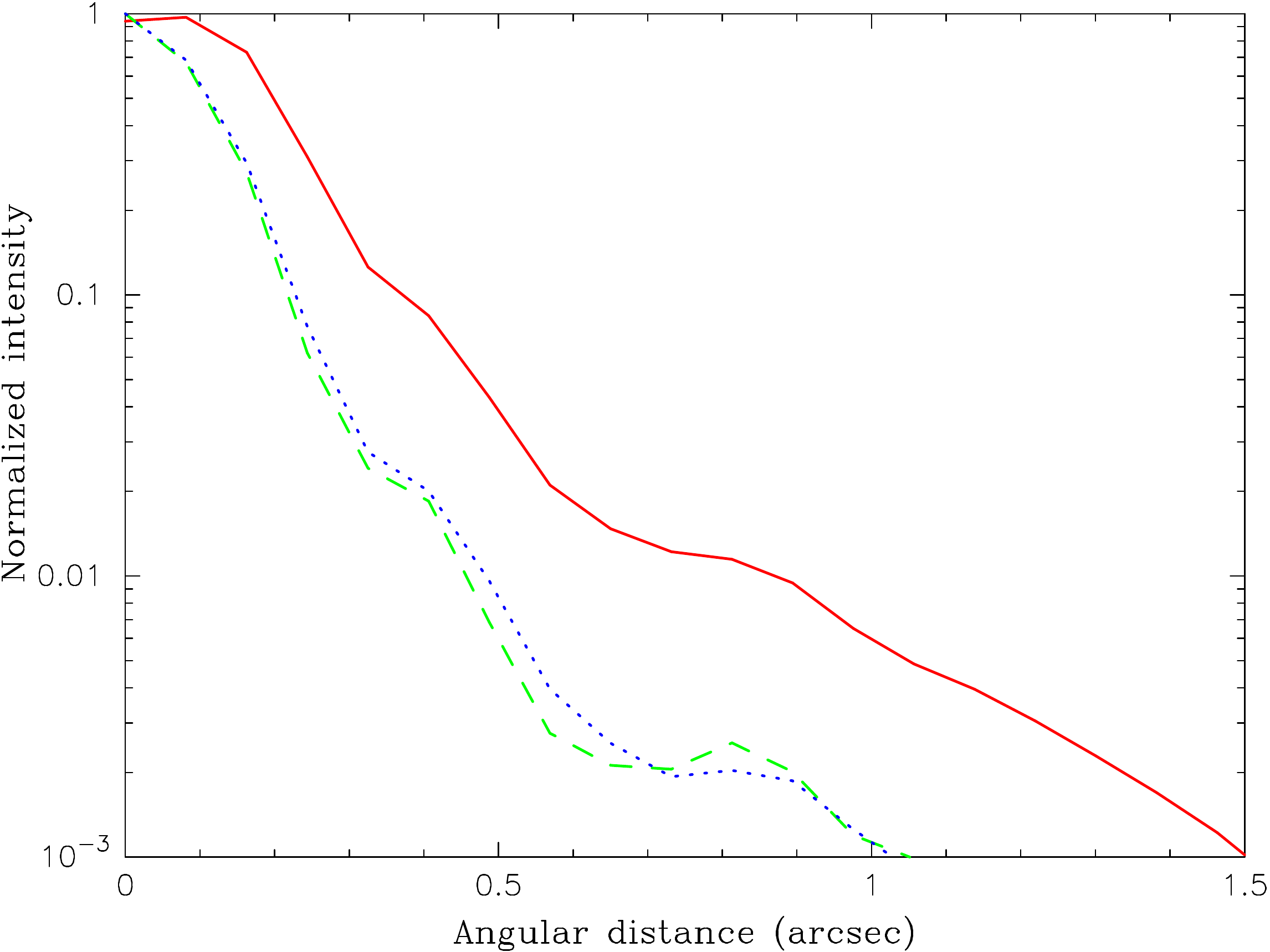}}}
\caption{
Azimuthally averaged intensity profile of \lpup\ (red solid line) and the PSF 
references $\varepsilon$~Cru (green dashed line) and S~Ori (blue dotted line).  
The central region of \lpup\ within 0\farcs1 is saturated. 
}
\label{l2pup_midi_acq_1Davg}
\end{figure}

Figure~\ref{l2pup_midi_acq} shows the 8.7~\micron\ images of \lpup\ and the 
PSF reference stars extracted from the MIDI acquisition data. 
The image of the first PSF reference $\varepsilon$~Cru 
(Fig.~\ref{l2pup_midi_acq}c) shows noticeable residual of the background 
subtraction, particularly near the edge of the field of view 
(large arcs near the top 
and bottom of the image).  The image of the second PSF reference star S~Ori 
is much better, and as Fig.~\ref{l2pup_midi_acq_1Davg} shows, 
the azimuthally averaged intensity profiles of two PSF reference stars agree, 
given the poorer quality of the $\varepsilon$~Cru image.  
The FWHM of the intensity profiles of the PSF references indicates a 
a spatial resolution of 220~mas, which corresponds to the diffraction limit 
at 8.7~\micron. 
The image of \lpup\ (Fig.~\ref{l2pup_midi_acq}a) as well as the azimuthally 
averaged intensity profile (Fig.~\ref{l2pup_midi_acq_1Davg}) clearly shows 
an envelope much more extended than the PSF references.  
The envelope extends to an angular radius of $\sim$1\farcs5 at the 0.1\% 
intensity of the (saturated) central peak. 
Figure~\ref{l2pup_midi_acq}b shows the image of \lpup\ with an intensity 
cutoff of 2\% of the central peak, in the same manner as the 11.7~\micron\ 
image presented by Jura et al. (\cite{jura02}, see their Fig.~2). 
Their 11.7~\micron\ image 
taken February 2001 (i.e., about four years before the MIDI data) shows 
an envelope with a radius of $\sim$1\arcsec, while our MIDI acquisition image 
at 8.7~\micron\ is much more compact with a radius of $\sim$0\farcs5 
when shown with the same intensity cutoff\footnote{The central peak of 
the \lpup\ image is saturated. 
If it were not for the saturation, the image with the 2\% intensity cutoff 
would appear even more compact than shown in Fig.~\ref{l2pup_midi_acq}b.}.  
This is probably because we probe the inner region of the envelope at 
8.7~\micron\ than at 11.7~\micron.  For this reason, we cannot draw a 
conclusion about time variations in the envelope between February 2001 and 
December 2004. 
In marked contrast to the near-IR image presented above, the mid-IR image 
appears approximately spherical without a signature of the disk.  
We discuss this point in Sect.~\ref{subsect_discuss_disk}.

\section{Discussion}
\label{sect_discuss}

\subsection{Comparison with the radiative transfer model of K14}
\label{subsect_discuss_rtmodel}

To examine whether our aperture-synthesis image of \lpup\ is quantitatively 
consistent with the 2D radiative transfer (RT) disk model of K14, 
we compared the measured interferometric observables---visibilities and 
closure phases---and the observed image with those predicted by their 
disk model.  P.~Kervella and M.~Montarg\`es kindly 
provided the best-fit model of their paper.  From their model at 2.17~\micron, 
which is the closest wavelength to our observations, 
we computed the 
visibilities and closure phases at the $u\varv$ points of our NACO speckle 
and AMBER data.  

\begin{figure*}
\resizebox{\hsize}{!}{\rotatebox{-90}{\includegraphics{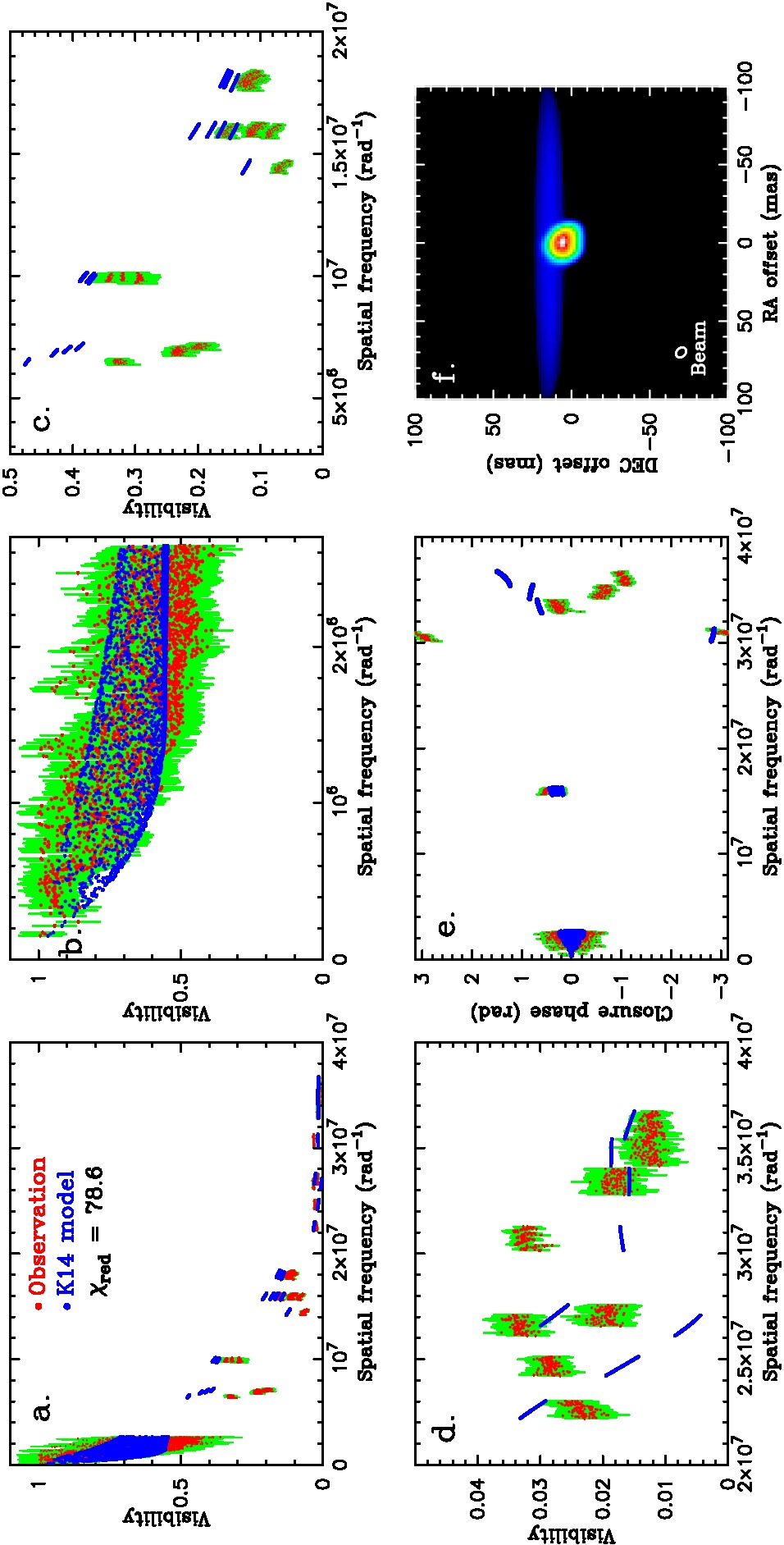}}}
\caption{
Comparison of our NACO speckle and AMBER observations with the 
RT model of K14 
(a color version of the figure is available in the electronic edition).  
In panels {\bf a}--{\bf e}, the observed data are plotted 
with the red (black in the printed edition) symbols with the error 
bars shown in green (light gray), while the model is 
represented by the blue (dark gray) symbols. 
{\bf a:} Visibilities. 
{\bf b:} Enlarged view of the visibilities obtained by 
speckle interferometry with NACO. 
{\bf c} and {\bf d:} Enlarged views of the visibilities obtained with AMBER 
in two different spatial frequency ranges. 
{\bf e:} Closure phases. 
{\bf f:} Model image in the same color scale as the reconstructed images in 
Fig.~\ref{reconst_images}. 
}
\label{l2pup_rtmodel}
\end{figure*}

Figure~\ref{l2pup_rtmodel} shows a comparison between the observed data and 
the RT model of K14.  
While the model reproduces the overall trend of the observed visibilities 
as a function of spatial frequency, the fit to the data is poor with 
the reduced $\chi^2$ = 78.6.  
The model image at 2.17~\micron\ convolved with the beam of our 
NACO+AMBER aperture-synthesis imaging (Fig.~\ref{l2pup_rtmodel}f) shows 
that the disk appears fainter and more inclined (i.e., closer to edge-on) 
compared to the observed image shown in 
Fig.~\ref{reconst_images}b (the color scale is the same for the model 
and observed images). 
The intensity of the disk in our aperture-synthesis image is $\sim$7\% 
of the central star, 
while it is $\sim$2\% in the convolved image of the RT model. 
This means that the flux contribution of the star is higher in the RT model 
than in the observed image, which also explains that the model visibilities are
higher than the observed data at spatial frequencies of 
$(7-20)\times10^6$~rad$^{-1}$. 

\begin{figure*}
\resizebox{\hsize}{!}{\rotatebox{-90}{\includegraphics{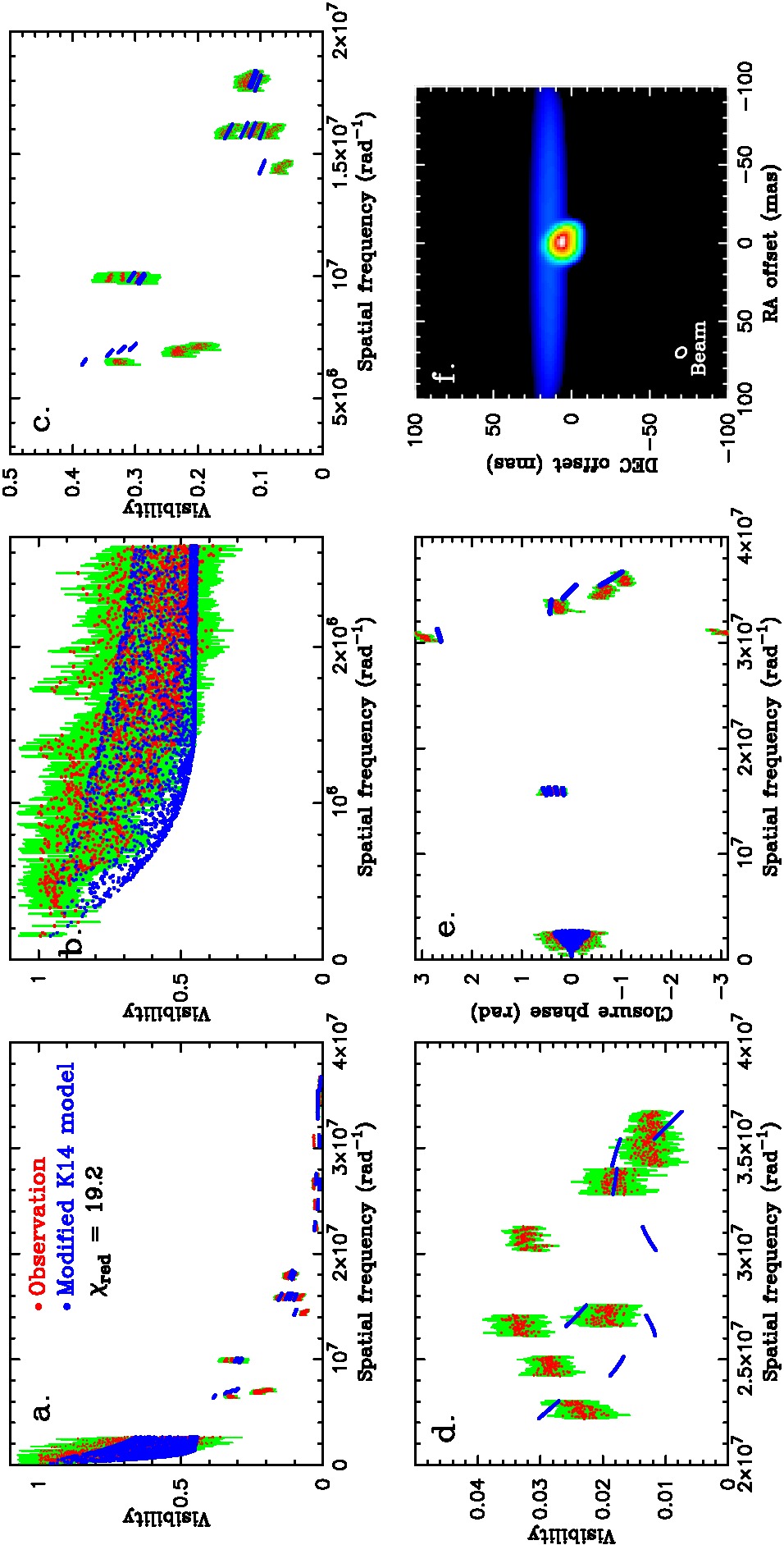}}}
\caption{
Comparison of our NACO speckle and AMBER observations with the 
modified RT model of K14, shown in the same manner as in 
Fig.~\ref{l2pup_rtmodel}. 
A color version of the figure is available in the electronic edition.  
}
\label{l2pup_rtmodel_modified}
\end{figure*}

It is possible 
that the flux contribution of the central star 
was higher at the time of the observations of K14 than for our observations. 
To examine this possibility, we changed the intensity of the star in the 
RT model image of K14.  
As Fig.~\ref{l2pup_rtmodel_modified} shows, the reduction of the stellar 
intensity by a factor of 0.6 can improve the match to the reduced $\chi^2$ = 
19.2. 
This result suggests that 
a time variation in the stellar flux by 0.55 magnitude at 2.2~\micron\ 
between our AMBER and NACO observations (December 2011 and March 2012) 
and the NACO observations of K14 (March 2013) can at least partially 
explain the disagreement between the RT model of K14 and our data. 
The amplitude of the $K$-band light curve of \lpup\ is $\sim$0.5 magnitude 
(Whitelock et al. \cite{whitelock00}).  Therefore, the variation in the 
stellar flux may be due to the pulsation of the central star. 
However, there is still disagreement in the visibilities observed at 
the longest baselines (Fig.~\ref{l2pup_rtmodel_modified}d), and 
the disk appears too inclined compared to the reconstructed image.  
This implies time variations in the structure of the disk, 
for example, inclination angle and/or the flaring angle of the disk.

\subsection{Binary companion}
\label{subsect_discuss_binary}

Kervella et al. (\cite{kervella15}) have recently found a companion at 
32.9~mas (2.1~AU) in west of the central star by adaptive 
optics polarimetric imaging in the visible with the VLT/SPHERE-ZIMPOL 
instrument. 
However, we do not see a signature of the companion 
in our NACO+AMBER image.  A possible reason might be that 
the companion is too faint at 2.2~\micron\ and/or our $u\varv$ coverage is 
too sparse in the E-W direction, and therefore, we might have failed 
to detect the signature of the companion in the visibility or closure phase. 
Alternatively, the companion might have been at a different position at the time 
of our observations.  Kervella et al. (\cite{kervella15}) estimated 
the orbital period to be 1.4--4.6~years. 
It is possible that the companion was behind the near side of the inner rim 
of the disk and was invisible at the time of our AMBER and NACO observations.

To set more quantitative constraints on the position and flux of the 
companion, we added a companion to the reconstructed image and examined the 
increase in the reduced $\chi^2$.  
The companion was placed at random positions within 
100~mas from the central star (corresponding to the AMBER field of 
view) and with a random flux contribution with respect to the star. 
We found out that the fit to the observed data is significantly poorer 
(reduced $\chi^2 > 2$) for a companion with a flux contribution higher 
than 2.4\% of the central star.  This sets an upper limit on the flux of 
the companion at 2.2~\micron.  On the other hand, we can set no constraints 
on the position of the companion.

The separation of the companion of 32.9~mas measured by Kervella et al. 
(\cite{kervella15}) corresponds to 3.8 stellar radii, given the angular 
diameter of 17.5~mas derived by K14. 
This implies that the primary red giant 
star may be filling the Roche lobe and may possibly show deformation, 
as seen in the Roche-lobe-filling systems $\beta$~Lyr and Algol = 
$\beta$~Per (Zhao et al. \cite{zhao08}; Baron et al. \cite{baron12}).  
We examined this possibility by introducing 
elliptical deformation to the central star in the initial 
model of the image reconstruction.  
If the major-axis of the star (E-W direction) is longer than 
the minor-axis (N-S direction) by more than 40\%, the reconstructed image 
cannot reproduce the observed data very well. 
The observed axis ratio of the Roche-lobe-filling star is $\sim$1.2 in the 
case of $\beta$~Lyr and Algol (Zhao et al. \cite{zhao08}; 
Baron et al. \cite{baron12}).  
Therefore, our $u\varv$ coverage is not sufficiently dense at long baselines 
to draw a definitive conclusion about the deformation. 
Aperture-synthesis imaging with a better $u\varv$ 
coverage is necessary to probe the shape of the primary star.

\subsection{Dimming event and the disk formation}
\label{subsect_discuss_disk}

As mentioned in Sect.~\ref{sect_intro}, 
\lpup\ started a dimming event around 1994.  If the dust clouds formed in 1994 
move away at a radial velocity of $\sim$3~\KMS\ 
(Winters et al. \cite{winters02}), 
they should have traveled a distance of $\sim$11~AU $\approx$ 170~mas 
by the time of our observations.  This is larger than the angular radius of 
the disk seen at 2.2~\micron,\ but comparable to the radius of the disk seen 
at 3--4~\micron\ (K14; L15). 
This suggests that the disk may have been formed by the dust clouds that 
have been ejected since the onset of the dimming event. 
The dust clouds ejected at the very beginning are perhaps already too far away 
from the star and, therefore, too cold to be detected at 2.2~\micron. 
We note, however, that this argument is based on the assumption that the dust 
clouds move away radially.  The 3.74~\micron\ images obtained by K14 and L15, 
the 4.05~\micron\ image of K14, and the visible image of Kervella et al. 
(\cite{kervella15}) show a spiral structure, which is presumably caused by 
the companion.  If the dust clouds move along the spiral structure, the 
radial distance traveled by the dust clouds is smaller than estimated above. 
Therefore, kinematical information of the spiral structure is necessary 
to confirm the connection between the formation of the disk and the 
onset of the dimming event.

While the disk seen in the near-IR is oriented roughly in the E-W direction, 
our mid-IR image at 8.7~\micron\ shown in Fig.~\ref{l2pup_midi_acq} as well as 
the 11.7~\micron\ and 17.9~\micron\ images presented in 
Jura et al. (\cite{jura02}) do not show a clear signature of the disk 
in this direction.  Our MIDI acquisition image appears approximately 
symmetric.  
The mid-IR images of Jura et al. (\cite{jura02}) are characterized 
by an elongation at a position angle of $\sim$135\degr\ and a distinct 
blob at 225\degr.  This makes the object appear slightly elongated 
in the E-W direction with a size of $\sim \!\! 2\farcs5 \times 2\farcs0$, 
but the elongation is much less pronounced than seen in the near-IR. Their 
spatial resolution is 0\farcs47 and 0\farcs49 at 11.7 and 17.9~\micron, 
respectively, 
which is significantly smaller than the angular size of the object, 
and therefore, the lower spatial resolution cannot fully account for 
the difference in the morphology.  
This means that the disk only exists in the innermost region of the 
circumstellar environment, while the outer region is approximately 
spherical at least at the time of the observations of Jura et al. 
(\cite{jura02}) and the MIDI observations.

The mid-IR imaging observations of Jura et al. (\cite{jura02}) took place 
in February 2001, while the MIDI data were obtained on December 2004. 
The dust clouds ejected at the onset of the dimming event (around 1994) 
should have traveled an angular distance of $\sim$70~mas 
(disk diameter of $\sim$140~mas) and $\sim$100~mas (disk diameter of 
$\sim$200~mas) at the time of Jura et al.'s observations and the MIDI 
observations, respectively. 
These angular displacements are smaller or just comparable to the 
angular resolution of these mid-IR observations, 
and therefore may have been too small to detect. 
Furthermore, while we assumed that the dust clouds move away radially, 
they may actually move along the spiral structure, in which case 
the radial distance would be smaller.  This might also have been a reason 
for the non-detection of the disk in the mid-IR observations. 
The circumstellar envelope imaged by these mid-IR observations most likely 
represents a nearly spherical mass loss that was already present 
before the formation of the disk.  

If we assume that the dust clouds move away radially, 
the dust clouds ejected at the beginning of the dimming 
event should have reached an angular distance of $\sim$200~mas by now 
(i.e., the disk diameter of $\sim$400~mas), which is resolvable with 
mid-IR instruments of the current 8--10~m telescopes.  
If the dust clouds move along the spiral structure, 
the angular size of the envelope should be smaller. 
Therefore, new mid-IR imaging would be interesting to examine whether or 
not the morphology of the circumstellar envelope on larger spatial scales 
has changed due to the formation of the disk and whether the measured size 
is consistent with the radial motion of the dust clouds or not.

\section{Concluding remarks}
\label{sect_concl}

We have obtained the aperture-synthesis image of both the disk and 
the central star of \lpup\ by combining the VLT/NACO speckle data and 
VLTI/AMBER long-baseline interferometric data.  The obtained image shows 
that the disk has a size of $\sim \!\!$ 180 $\times$ 50~mas and the 
southern half of the central star is severely obscured by the equatorial 
dust lane of the nearly edge-on disk.  
This agrees with the recent AO and aperture-masking images and modeling. 
The observed size of the disk is consistent with the angular distance 
that the dust clouds formed at the beginning of the dimming event 
($\sim$1994) should have traveled by the time of our observations, 
if we assume that the dust clouds move away radially. 
This lends support to the hypothesis that the formation of the dust disk is responsible 
for the dimming event. 
However, 
the kinematical information of the spiral structure is important for 
establishing the connection between the formation of the disk and the 
dimming event.  ALMA observations of molecular lines are ideal for 
this goal.

The mid-IR image extracted from the MIDI data taken about seven years 
before our AMBER and NACO observations shows an approximately spherical 
envelope, which most likely results from the mass loss occurring 
before the disk formation. 
High-resolution mid-IR imaging is useful for studying possible 
morphological changes in the circumstellar environment on larger spatial 
scales since the onset of the dimming event.

\begin{acknowledgement}
We thank the ESO Paranal team for supporting our AMBER and NACO observations.  
We are also grateful to the ESO's Director General Tim de Zeeuw for 
allocating our NACO observations in the Director's Discretionary Time 
and to Pierre Kervella and Miguel Montarg\`es for providing us with their 
radiative transfer model images. 
\end{acknowledgement}

\appendix

\section{Removal of detector artifacts from MIDI acquisition images}
\label{appendix_midi}

\begin{figure*}
\sidecaption
\resizebox{12cm}{!}{\rotatebox{0}{\includegraphics{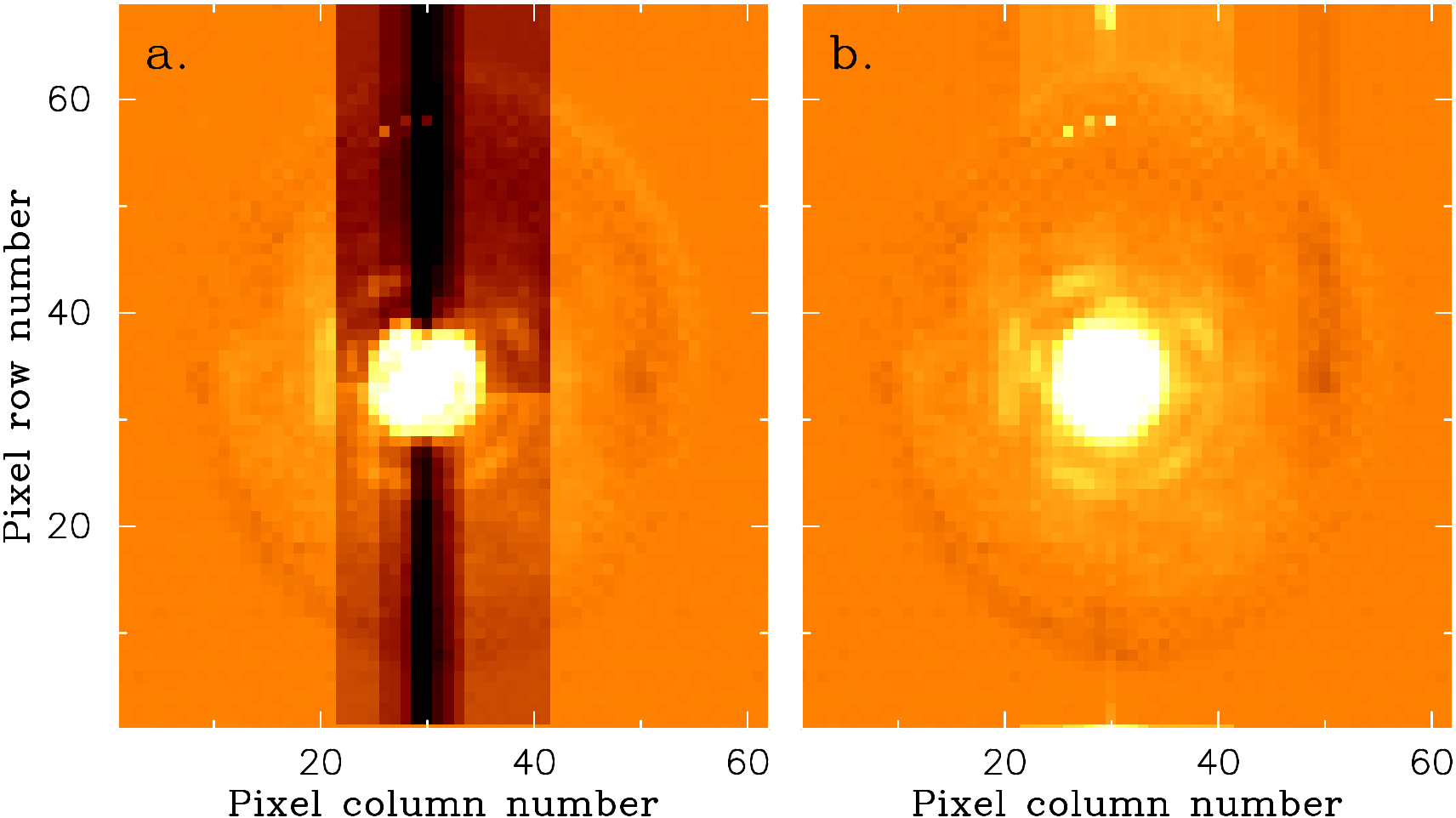}}}
\caption{
Removal of the detector artifacts in the MIDI acquisition images of \lpup. 
{\bf a:} One of the sky-subtracted frame of \lpup, showing vertical stripes 
in the columns near the center. 
{\bf b:} Same frame after the removal of the vertical stripes. 
The color scale in the diffraction core is saturated to clearly show 
the stripes. 
}
\label{l2pup_destriping}
\end{figure*}

As Fig.~\ref{l2pup_destriping}a shows, noticeable vertical stripes are 
present in the sky-subtracted frames of the MIDI acquisition data of \lpup. 
These detector artifacts presumably result from the high brightness of 
\lpup\ (the central region within 0\farcs1 is saturated).  No vertical 
stripes appear in the images of the PSF reference stars.  
The intensity of the vertical stripes is different in the upper and lower 
regions with respect to the star, and it therefore is necessary to remove 
the stripes in the upper and lower regions separately.  
In each region, we removed the stripes as follows. 
First, for each column affected by the vertical stripes, we computed the 
median of the pixel values from the rows sufficiently far away from the star.  
Then we subtracted this median from all pixels in the column.  We carried out 
this procedure for all columns that were affected by the stripes. 
Figure~\ref{l2pup_destriping} demonstrates that the image is nearly free from 
the stripes after this procedure.

\section{Image reconstruction}
\label{appendix_reconst}

The observed visibilities plotted in Fig.~\ref{reconst_interf}a show 
a steep drop at low spatial frequencies $\la \! 3\times10^6$~rad$^{-1}$ 
(= baselines shorter than 6~m), which suggests a very extended component.  
The visibilities observed at 
spatial frequencies from $\sim \!\! 6\times10^6$~\PERRAD\ to 
$10^7$~\PERRAD\ (baselines from 13 to 22~m) may 
appear reminiscent of the first and the second visibility lobe expected 
from a uniform disk or limb-darkened disk.  
However, the visibility from a uniform disk without an extended component 
is 0.13 (or lower for a limb-darkened disk) 
in the extrema of the second visibility lobe, 
much lower than the observed values of $\sim$0.3.  
With an extended component as revealed by the speckle data, the visibility 
would be even lower.  
We first attempted to explain these observed data using geometrical models. 
While simple geometrical models may not fit the data completely, they are 
useful for characterizing the approximate geometry of the object 
and can also be used as an initial model for the image reconstruction.

\subsection{Geometrical model: uniform disk + elliptical Gaussian}
\label{subsect_gaussian_model}

\begin{figure*}
\resizebox{\hsize}{!}{\rotatebox{-90}{\includegraphics{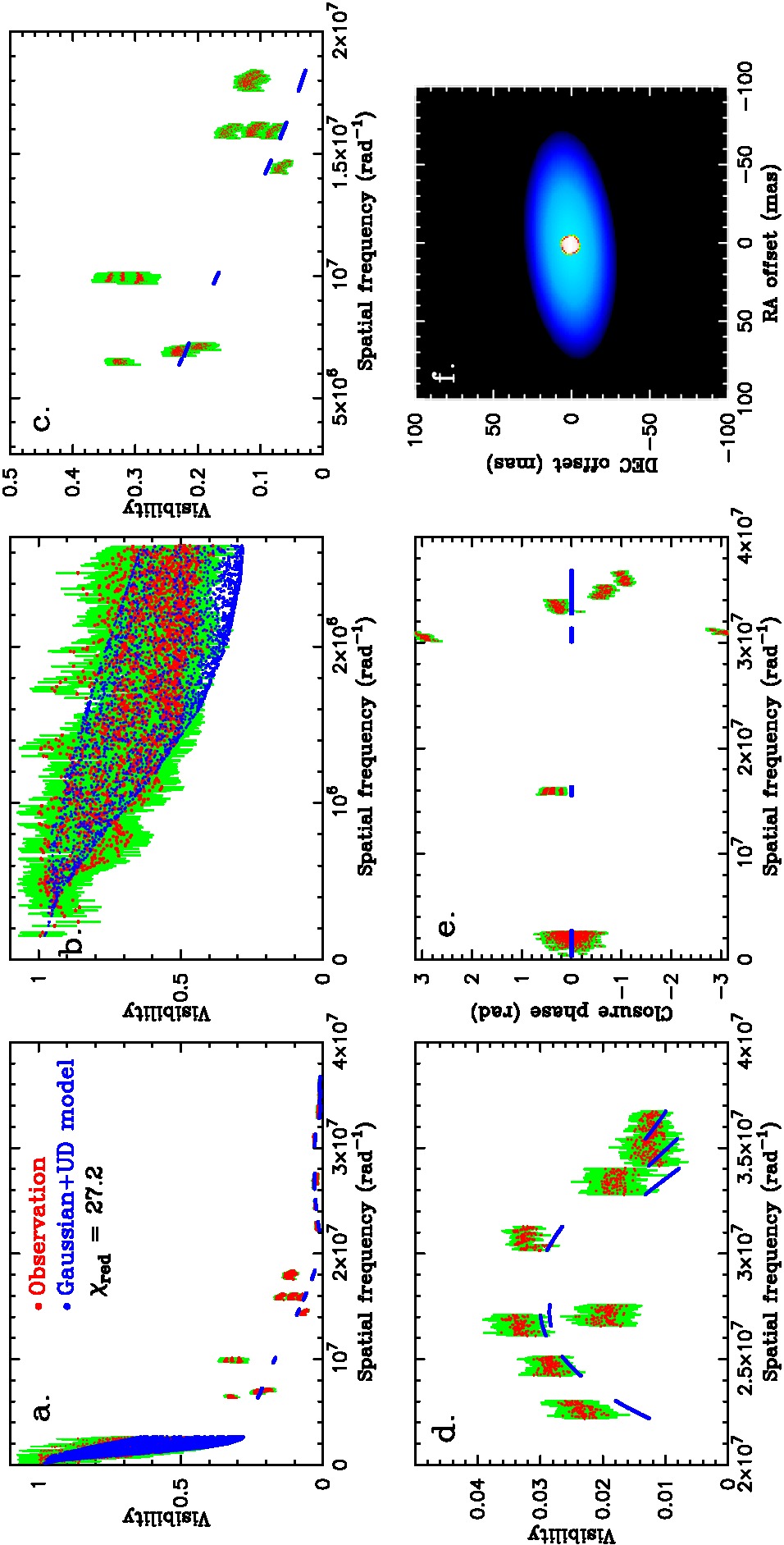}}}
\caption{
Best-fit model consisting of a uniform-disk central star and an elliptical 
Gaussian shown in the same manner as Fig.~\ref{l2pup_rtmodel}. 
}
\label{gauss_model}
\end{figure*}

We tried to fit the data with a geometrical model consisting 
of a uniform-disk-like central star and an elliptical Gaussian. 
The free parameters are the uniform-disk diameter of the central star 
(\PHISTAR), 
the fractional flux contribution of the central star $f_{\star}$, 
the widths of the elliptical Gaussian along the major and minor axes 
\SIGMAEW\ and \SIGMANS\ (the elliptical Gaussian is given by 
$e^{-((x/\SIGMAEW)^2 +(y/\SIGMANS)^2)}$), and the position angle of its 
major axis PA (measured from North to East).  
We searched for the best-fit model by varying \PHISTAR\ = 8 ... 22 (mas) 
with $\Delta \PHISTAR $ = 2 (mas), 
$f_{\star}$ = 0.1 ... 0.7 with $\Delta f_{\star}$ = 0.05, 
\SIGMAEW\ = 30 ... 100 (mas) with $\Delta \SIGMAEW$ = 10 (mas), 
\SIGMANS\ = 10 ... 50 (mas) with $\Delta \SIGMANS$ = 10 (mas), 
and PA = 70\degr\ ... 100\degr\ with $\Delta {\rm PA}$ = 5\degr\ 
(the PA was limited to this range from the elongation of the 
image reconstructed from the speckle data alone). 
The best-fit model, which is plotted in Fig.~\ref{gauss_model}, is 
characterized by \PHISTAR\ = 12~mas, \FSTAR\ = 0.3, \SIGMAEW\ = 50~mas, 
\SIGMANS\ = 20~mas, and PA = 95\degr\ with a reduced $\chi^2$ of 27.2. 
Figure~\ref{gauss_model} reveals that the best-fit model cannot reproduce 
the observed visibilities at spatial frequencies of 
$(0.6-2.0) \times 10^7$~\PERRAD\ (baseline length = 13--44~m). 
Most of the observed visibilities at these spatial frequencies are noticeably 
higher than predicted by the model.  
This means that there is some sharp structure that is not seen in the 
smooth Gaussian model.

\subsection{Geometrical model: uniform disk + elliptical ring}
\label{subsect_ring_model}

\begin{figure*}
\resizebox{\hsize}{!}{\rotatebox{-90}{\includegraphics{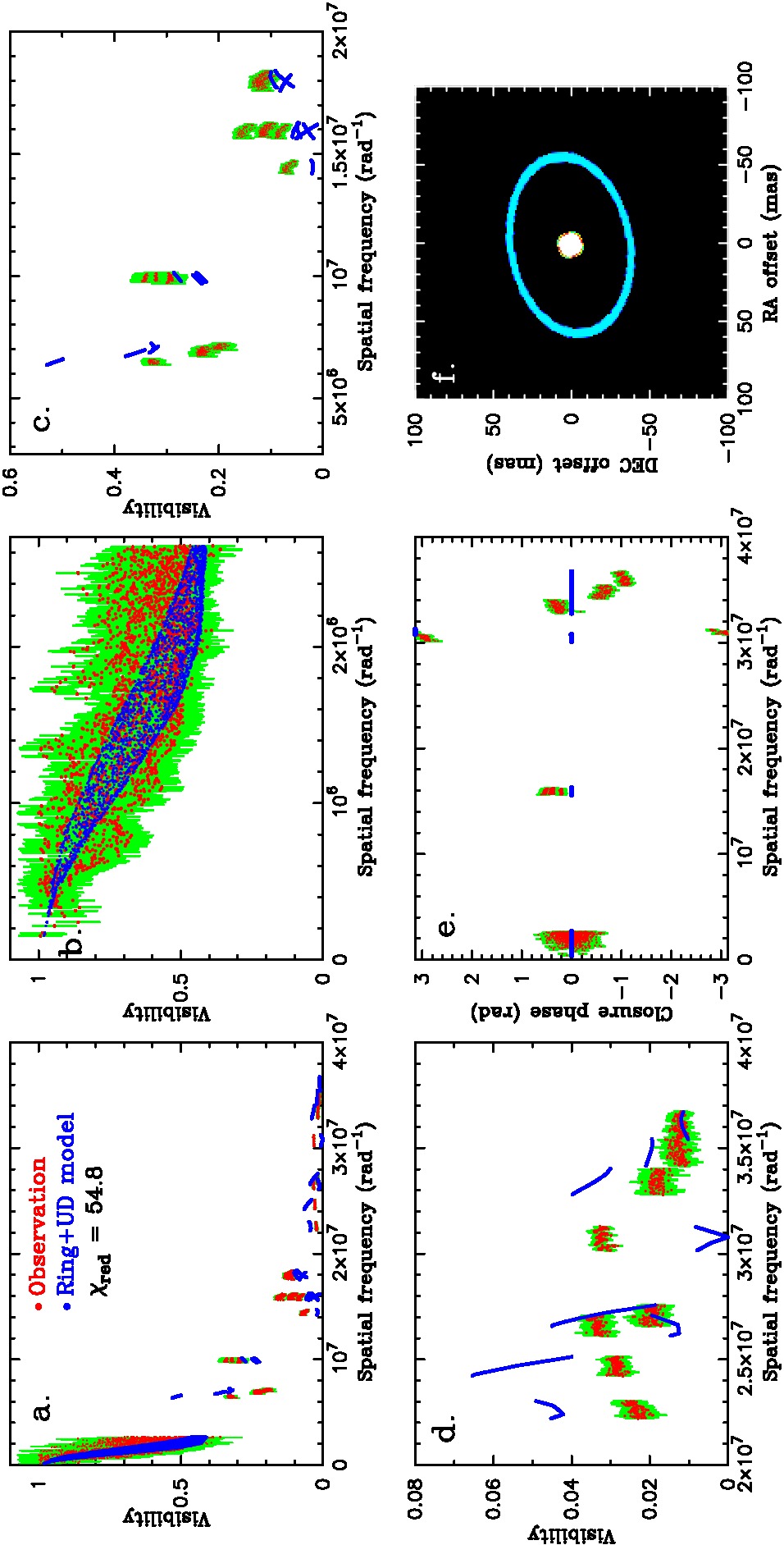}}}
\caption{
Best-fit model consisting of a uniform-disk central star and an elliptical 
ring shown in the same manner as Fig.~\ref{l2pup_rtmodel}. 
}
\label{ring_model}
\end{figure*}

We then tried a model consisting of the 
uniform-disk-like central star and an elliptical ring, because a ring gives 
rise to visibilities much higher than a uniform disk or limb-darkened disk at 
long baselines.  
The free parameters are the uniform-disk diameter of the 
central star \PHISTAR, the fractional flux contribution of the central star 
\FSTAR, the semi-major and semi-minor axes of the elliptical ring 
(\RRINGEW\ and \RRINGNS, respectively), and its position angle PA. 
The width of the ring was set to be 10\% of its radius at each position angle. 
We searched for the best-fit model by varying 
\PHISTAR\ = 8 ... 22 (mas) with $\Delta \PHISTAR $ = 2 (mas), 
$f_{\star}$ = 0.1 ... 0.6 with $\Delta f_{\star}$ = 0.05, 
\RRINGEW\ = 30 ... 100 (mas) with $\Delta \RRINGEW$ = 10 (mas), 
\RRINGNS\ = 10 ... 50 (mas) with $\Delta \RRINGEW$ = 10 (mas), 
and PA = 70\degr\ ... 100\degr\ with $\Delta {\rm PA}$ = 5\degr.  
Figure~\ref{ring_model} shows a comparison of the best-fit ring model with 
the observed data.  This model is characterized by \PHISTAR\ = 16~mas, 
\FSTAR\ = 0.6, \RRINGEW\ = 60~mas, \RRINGNS\ = 40~mas, and PA = 95\degr\ with 
the reduced $\chi^2$ = 54.8.  
While the visibilities at $(0.6-1)\times10^7$~\PERRAD\ from this model 
is as high as or even higher than the observed data, the fit to the 
speckle data and the data at longer baselines is worse than the above 
star + Gaussian model as shown by the worse reduced $\chi^2$.

\subsection{Geometrical model: obscured uniform disk + elliptical Gaussian}
\label{subsect_obscured_gauss_model}

We found out that the observed data are much better reproduced 
if the southern half of the aforementioned uniform-disk-like central star + 
elliptical Gaussian model is obscured.  
In this model, the intensity of the uniform-disk star + elliptical Gaussian 
model $I_0(x,y)$ ($x$ and $y$ are the coordinates on the sky with the 
origin at the central star) is modified as follows: 
\[
I(x,y) = I_0(x,y) \times \frac{1}{e^{-y/\varepsilon}+1}, 
\]
where $\varepsilon$ is a parameter to smoothen the obscuration edge. 
We set $\varepsilon$ to be 0.2, which decreases the intensity to 
zero over $\sim$2~mas in the $y$ direction.  
The free parameters are the uniform-disk diameter of the central star, 
the fractional flux contribution of the central star $f_{\star}$, 
the widths of the elliptical Gaussian along the major and minor axes 
\SIGMAEW\ and \SIGMANS, 
and the position angle of its major axis PA.  
We searched for the best-fit model by varying \PHISTAR\ = 8 ... 22 (mas) 
with $\Delta \PHISTAR $ = 2 (mas), 
$f_{\star}$ = 0.2 ... 0.6 with $\Delta f_{\star}$ = 0.05, 
\SIGMAEW\ = 30 ... 100 (mas) with $\Delta \SIGMAEW$ = 10 (mas), 
\SIGMANS\ = 10 ... 50 (mas) with $\Delta \SIGMANS$ = 10 (mas), 
and PA = 70\degr\ ... 100\degr\ with $\Delta {\rm PA}$ = 5\degr. 

\begin{figure*}
\resizebox{\hsize}{!}{\rotatebox{-90}{\includegraphics{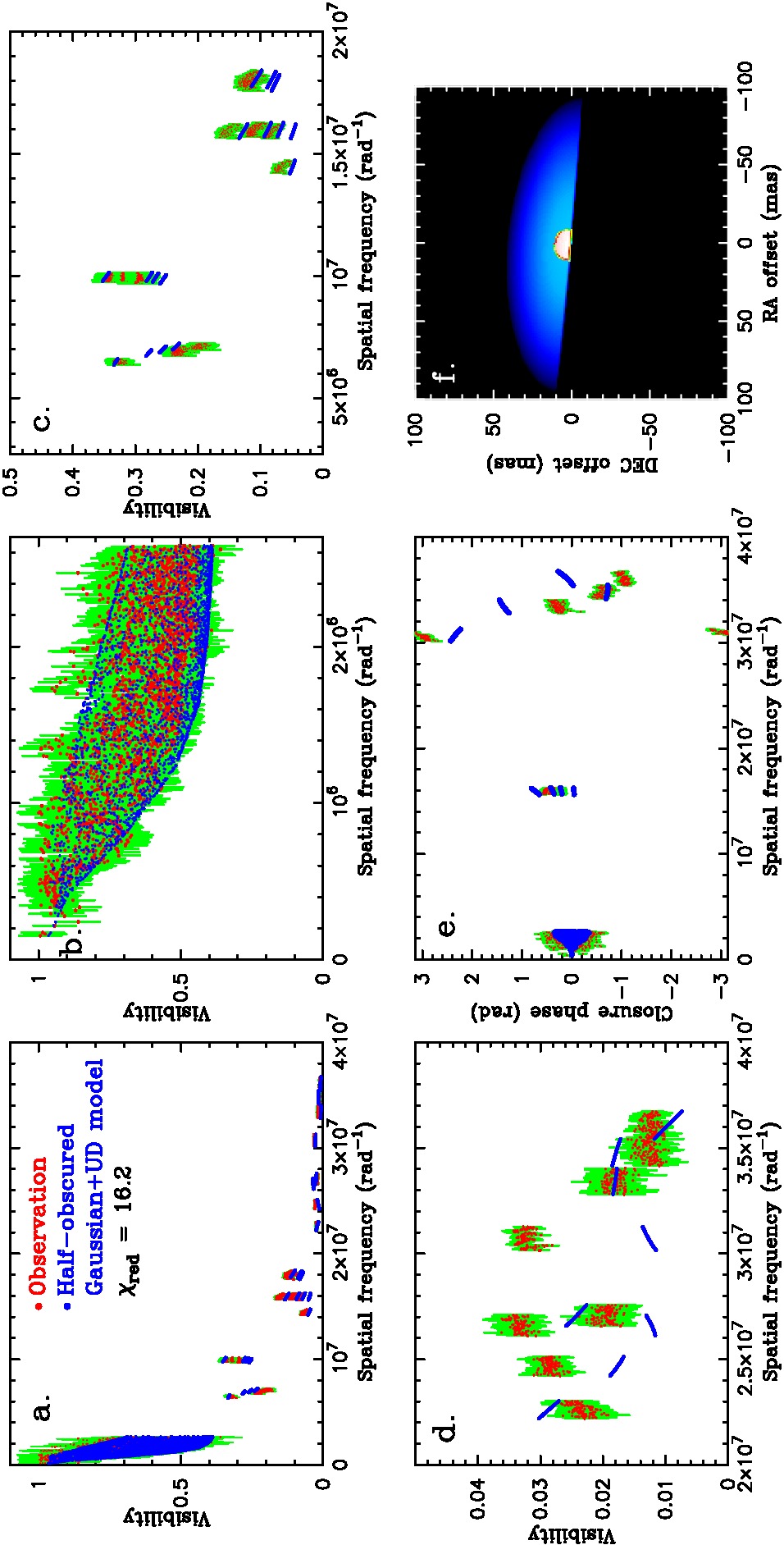}}}
\caption{
Best-fit model consisting of a half-obscured uniform-disk central star 
and an elliptical Gaussian shown in the same manner as 
Fig.~\ref{l2pup_rtmodel}. 
}
\label{obscured_gauss_model}
\end{figure*}

The best-fit model is characterized by \PHISTAR\ = 20~mas, 
\FSTAR\ = 0.45, \SIGMAEW\ = 70~mas, \SIGMANS\ = 30~mas, and PA = 85\degr. 
As Fig.~\ref{obscured_gauss_model} shows, this model can reproduce the 
observed visibilities much better than the above two models. 
The uncertainties in \FSTAR, \SIGMAEW, \SIGMANS, and PA are 
$\pm0.05$, $\pm10$~mas, 
$\pm10$~mas, and $\pm5$~\degr.  The uniform-disk diameter of the central 
star is in the range between 18 and 20~mas.  This agrees with 
the 17.5~mas derived by K14, given the difference in the data and the 
model used by them and us. 
The reduced $\chi^2$ of this model is 16.2, which is much better than the 
star + Gaussian or star + ring models, but still much larger than 1.  
This is because the fit to the visibilities observed at the longest 
baselines (spatial frequencies higher than $\sim \!\! 2\times 10^7$~\PERRAD) 
and to the observed closure phases are not satisfactory.  
However, this disagreement can be due to small-scale structures that are 
not included in our geometrical model, but can be modeled by the image 
reconstruction. 

\subsection{MiRA parameters}
\label{subsect_mira}

We used the best-fit half-obscured star + elliptical Gaussian model 
as the initial model.  The regularization scheme of the maximum entropy 
method was adopted, with the prior being the Gaussian with the same 
widths as the initial model.  
The degree of regularization was set to $\mu = 10^3$ 
(see Thi\'ebaut \cite{thiebaut08} for details of the regularization 
scheme and the definition of $\mu$).  
We also reconstructed the image with the total variation regularization 
scheme, but the image shows no noticeable differences.

\end{document}